\documentclass[letterpaper,12pt, onecolumn]{IEEEtran}

\usepackage{subfig} % Subfigures

\usepackage[total={6in,9in},top=1in,left=1.25in,bindingoffset=0in]{geometry}
\usepackage[normalem]{ulem} % Underlines that allow line breaks
\usepackage{pifont} % arrows / hands / pens / scisors
\usepackage[usenames]{color} % colors
\usepackage{amsmath} % Extended math notation
\usepackage{amsfonts} % Extra math fonts
\usepackage{array} % For ragged text on tables
\usepackage{graphicx} % Include figures
\usepackage[english]{babel} % Hypenation

\usepackage{amsmath} % Extended math notation
\usepackage{amsfonts} % Extra math fonts
\usepackage{graphicx} %plots in Latex
\usepackage{amsthm}
\usepackage{geometry}

\usepackage{pgf}
\usepackage{tikz}
\usepackage{pgfplots}
\usetikzlibrary{arrows,automata}
\usetikzlibrary{positioning}
\usetikzlibrary{backgrounds}
\usetikzlibrary{matrix}
\usetikzlibrary{decorations.pathmorphing}
\usetikzlibrary{fit}

\newtheorem{lemm}{Lemma}
\newtheorem{prop}{Proposition}
\newtheorem{theo}[lemm]{Theorem}

\newenvironment{demo}{\noindent {\indent{\em Proof: \ }}\ }{\qed}

\title{A realistic distributed storage system that minimizes data storage and
repair bandwidth.
\footnote[1]{We want to thank professor Alexandros G. Dimakis for
his suggestions and
contributions to this paper. This work has been partially supported by the
Spanish MICINN grant TIN2010-17358, the Spanish Ministerio de Educaci\'on FPU
grant AP2009-4729 and the Catalan AGAUR grant 2009SGR1224.}}

\author{ \vspace{0.25cm}Bernat Gast\'on, Jaume Pujol, and Merc\`e Villanueva\\
\small Department of Information and Communications Engineering \\
\small Universitat Aut\`onoma de Barcelona \\
\small Cerdanyola del Vall\`{e}s (Barcelona), Spain \\
\small \texttt{\{Bernat.Gaston $|$ Jaume.Pujol $|$
Merce.Villanueva \}@uab.cat} \\
}
\begin{document}
\maketitle

\begin{abstract}
In a realistic distributed storage environment, storage nodes are usually
placed in racks, a metallic support designed to accommodate electronic
equipment. It is known that the communication (bandwidth) cost between nodes
within a rack is much lower than the communication (bandwidth) cost  between
nodes within different racks.

In this paper, a new model, where the storage nodes are placed in two racks,
is proposed and analyzed. In this model, the storage nodes have different repair costs to
repair a node depending on the rack where they are placed. A threshold
function, which minimizes the amount of stored data per node and the bandwidth
needed to regenerate a failed node, is shown. This threshold function
generalizes the threshold function from previous distributed storage models. The
tradeoff curve obtained from this threshold function is compared with the ones
obtained from the previous models, and it is shown that this new model
outperforms the previous ones in terms of repair cost.
\end{abstract}

\section{Introduction}

In a distributed storage environment, where the data is
placed in nodes connected through a network, it is likely that one of these
nodes fails. It is known that the use of erasure coding improves the fault
tolerance and minimizes the amount of stored data \cite{Rod01}, \cite{We01}.
Moreover, the use of regenerating codes  not only makes the most of the erasure
coding improvements, but also minimizes the amount of data needed to regenerate
a failed node \cite{Di01}.

In realistic distributed storage environments for example a storage cloud, the
data is placed in storage devices which are connected through a network. These
storage devices are usually organized in a rack, a metallic support designed to
accommodate electronic equipment. The communication (bandwidth) cost between
nodes within a rack is much lower than the communication (bandwidth) cost
between nodes within different racks.

In \cite{Di01}, an optimal tradeoff between the amount of stored data per node
and the repair bandwidth needed to regenerate a failed node (repair bandwidth)
in a distributed storage environment was claimed. This tradeoff was
proved by using the mincut on information flow graphs, and it can be represented as a
curve, where the two extremal points of the curve are called the Minimum Storage
Regenerating (MSR) point and the Minimum Bandwidth Regenerating (MBR) point.

In \cite{Ak01}, another model, where there is a static classification of ``cheap bandwidth''
and ``expensive bandwidth'' storage nodes, was introduced. However, this
classification is not based on racks, because the nodes in the expensive set are
always expensive in terms of repair cost, regardless of the failed node.

This paper is organized as follows. In Section \ref{sec:1}, we analyze previous
distributed storage models.
In Section \ref{rackModel}, we provide a new model, where the storage nodes are
placed in two racks. We also provide a general threshold function and we specify
the MBR and MSR points in this model.
In Section \ref{sec:3}, we analyze the results of this new model compared to the
previous ones. Finally, in Section \ref{sec:4}, we expose the conclusions of
this study.

\section{Previous models}
\label{sec:1}

In this section, we will describe the previous distributed storage models: the
basic model and the static cost model introduced in \cite{Di01} and \cite{Ak01}, respectively.
\subsection{Basic model}
\label{FirstModel}

In \cite{Di01}, \textit{Dimakis et al.} introduced a first distributed storage
model, where there is the same repair cost between any two storage nodes.
Moreover, the fundamental tradeoff between the amount of stored data per node
and the repair bandwidth was given from analyzing the mincut of an information
flow graph.

Let $C$ be a $[n,k,d]$ regenerating code composed by $n$ storage nodes, each
one storing $\alpha$ data units, and such that any $k$ of these $n$ storage
nodes contain enough information to recover the file. In order to be able to
recover a file of size $M$, it is necessary that $\alpha k \ge M$.
When one node fails, $d$ of the remaining $n-1$ storage nodes send $\beta$ data
units to the new node which will replace the failed one. The new node is called
newcomer, and the set of nodes sending data to the newcomer are called helper
nodes. The total amount of bandwidth used per node regeneration is $\gamma= d
\beta$.

\begin{figure}
\centering
\begin{tikzpicture}[shorten >=1pt,->]
  \tikzstyle{vertix}=[circle,fill=black!25,minimum size=18pt,inner sep=0pt,
node distance = 0.8cm,font=\tiny]
  \tikzstyle{invi}=[circle]
  \tikzstyle{background}=[rectangle, fill=gray!10, inner sep=0.2cm,rounded
corners=5mm]

  \node[vertix] (s) {$S$};

  \node[vertix, right=1cm of s] (vin_2)  {$v_{in}^2$};
  \node[vertix, below of=vin_2] (vin_3)  {$v_{in}^3$};
  \node[vertix, above of=vin_2] (vin_1)  {$v_{in}^1$};
  \node[vertix, below of=vin_3] (vin_4)  {$v_{in}^4$};

 \foreach \to in {1,2}
   {\path (s) edge[bend left=20,font=\tiny] node[anchor=south,above]{$\infty$}
(vin_\to);}
 \foreach \to in {3,4}
   {\path (s) edge[bend right=20,font=\tiny]  node[anchor=south,above]{$\infty$}
 (vin_\to);}

  \node[vertix, right of=vin_1] (vout_1)  {$v_{out}^1$};
  \node[vertix, right of=vin_2] (vout_2)  {$v_{out}^2$};
  \node[vertix, right of=vin_3] (vout_3)  {$v_{out}^3$};
  \node[vertix, right of=vin_4] (vout_4)  {$v_{out}^4$};

  \node[vertix, right of=vout_4, node distance = 2cm] (vin_5)
{\scriptsize{$v_{in}^5$}};
  \node[vertix, right of=vin_5] (vout_5)
{\scriptsize{$v_{out}^5$}};

  \node[vertix, right of=vout_1, above of=vin_5, node distance = 2cm] (vin_6)
{\scriptsize{$v_{in}^6$}};
  \node[vertix, right of=vin_6] (vout_6)
{\scriptsize{$v_{out}^6$}};

  \node[vertix, right of=vout_6, node distance = 1.5cm] (DC)
{DC};

   \path (vout_5) edge[bend right=20,font=\tiny]
node[anchor=south,above]{$\infty$}
 (DC);
   \path (vout_6) edge[bend left=10,font=\tiny]
node[anchor=south,above]{$\infty$}
 (DC);

  \path[->, bend left=15,font=\tiny] (vout_2) edge
node[anchor=south,above]{$\beta$} (vin_5);
  \path[->, bend left=10,font=\tiny] (vout_3) edge
node[anchor=south,above]{$\beta$}(vin_5);
  \path[->, bend left=20,font=\tiny] (vout_1) edge
node[anchor=south,above]{$\beta$}(vin_5);

  \path[->, bend left=15,font=\tiny] (vout_1) edge
node[anchor=south,above]{$\beta$} (vin_6);
  \path[->, bend left=10,font=\tiny] (vout_2) edge
node[anchor=south,above]{$\beta$}(vin_6);
  \path[->, bend left=10,font=\tiny] (vout_5) edge
node[anchor=south,above]{$\beta$}(vin_6);

 \foreach \from/\to in {1,2,3,4,5,6}
  { \path[->,font=\tiny] (vin_\from) edge node[anchor=south] {$\alpha$}
(vout_\to); }

  \draw[-,color=red,thick] (1.2,-0.5) -- (2.7,-1.2);
  \draw[-,color=red,thick] (1.2,-1.2) -- (2.7,-0.5);

  \draw[-,color=red,thick] (1.2,-1.2) -- (2.7,-1.9);
  \draw[-,color=red,thick] (1.2,-1.9) -- (2.7,-1.2);
\end{tikzpicture}

\caption{Information flow graph corresponding to a $[4,2,3]$ regenerating code.
}
\label{example1}
\end{figure}
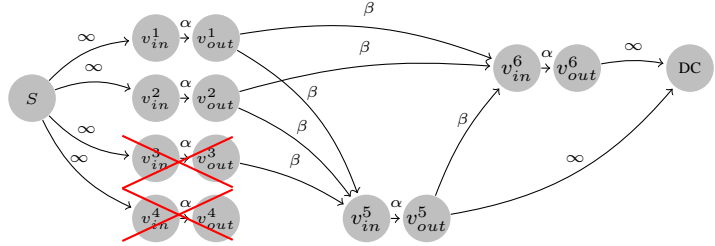

Let $s_i$, where $i = 1,\ldots, \infty$, be the $i$-th storage node. Let $G(V,E)$
be a weighted graph designed to represent the information flow. Then, $G$ is in
fact a directed acyclic graph, with a set of vertices $V$ and a set of arcs
$E$.
The set $V$ is composed by three kinds of vertices:
\begin{itemize}
 \item Source vertex $S$: there is only one source vertex in the graph, and
it represents the file to be stored.
 \item Data collector vertex $DC$: it represents the user who is allowed to
access the data in order to reconstruct the file.
 \item Storage node vertices $v_{in}^i$ and $v_{out}^i$: each storage node
$s_i$, where $i=1,\ldots, \infty$, is represented by one inner vertex $v_{in}^i$
and one outer vertex $v_{out}^i$.
Let $V_s \subset V$ be the set of all these storage node vertices.
\end{itemize}
In general, there is an arc $(v,w) \in E$ of weight $c$ from vertex
$v \in V$ to vertex $w \in V$ if vertex $v$ can send $c$ data units to
vertex $w$.

At the beginning of the life of a distributed storage environment, there is
a file to be stored in $n$ storage nodes $s_i$, $i=1,\ldots,n$. This
means that there is a source vertex $S$ with outdegree $n$ connected to
vertices $v_{in}^i$, $i=1,\ldots,n$. Since we want to analyze the information flow of
graph $G$ in terms of $\alpha$ and $\beta$, and these $n$ arcs are not
significant to find the mincut of $G$, their weight is set to infinite.
Each one of the storage nodes $s_i$, $i=1,\ldots,n$, stores
$\alpha$ data units. To represent this fact, each vertex $v_{in}^i$ is connected
to vertex $v_{out}^i$ with an arc of weight $\alpha$.

When the first storage node fails,
the newcomer node $s_{n+1}$ connects to $d$ existing storage nodes sending,
each one of them, $\beta$ data units. So, there is one arc from $v_{out}^i$,
$i=1,\ldots,n$, to $v_{in}^{n+1}$ with weight $\beta$ if $s_i$ sends $\beta$
data
units to $s_{n+1}$ in the regenerating process. The new vertex $v_{in}^{n+1}$ is
also connected to its associated $v_{out}^{n+1}$ with an arc of weight $\alpha$.
This process can be repeated for every failed node. Let the
new storage nodes (newcomers) be $s_j$, where $j=n+1,\ldots,\infty$.

Finally, after some failures, a data collector wants to reconstruct the file.
Therefore, a vertex $DC$ is also added to the graph. There is one arc from
vertex $v_{out}^i$ to $DC$ if the data collector connects to the storage node
$s_i$. Note that if $s_i$ has been replaced by $s_j$, this means that the
vertex $DC$ can not connect to $v_{out}^i$, but it can connect to
$v_{out}^j$. The vertex
$DC$ has indegree $k$ and each arc has weight infinite, because
they have no relevance in finding the mincut of $G$.

If the mincut from vertex $S$ to $DC$ achieves $\mbox{mincut}(S,DC)\ge M$, it means that the data
collector can reconstruct the file, since there is enough information flow
from the source to the data collector.
In fact, the data collector can connect to any $k$ nodes, so $\min(\mbox{mincut}(S,DC))\ge
M$, which is achieved when the data collector connects to $k$ storage nodes that
have already been replaced by a newcomer \cite{Di01}. From this scenario, the
mincut is computed and lower bounds on
the parameters $\alpha$ and $\gamma$ are given.
Let $\alpha^*(d,\gamma)$ be the threshold function, which is the function that
minimizes $\alpha$. As $\alpha \ge \alpha^*(d,\gamma)$, if $\alpha^*(d,\gamma)$
can be achieved $\alpha$ is possible too.

Figure \ref{example1} illustrates the information flow graph $G$ associated
to a $[4,2,3]$ regenerating code. Note that $\mbox{mincut}(S,DC) = \min
\left\lbrace 3\beta, \alpha \right\rbrace + \min \left\lbrace 2\beta, \alpha
\right\rbrace$. For a general information flow graph, $\mbox{mincut}(S,DC)
\ge \sum_{i=0}^{k-1} \min \left\lbrace (d-i)\beta,
\alpha \right\rbrace \ge M$, which after an optimization process leads to

\begin{equation}
\label{thresholdDim}
 \alpha^*(d,\gamma) = \left\{ \begin{array}{lc}
             \frac{M}{k}, & \gamma \in [f(0), +\infty) \\
             \\ \frac{M-g(i)\gamma}{k-i}, & \gamma \in [f(i),f(i-1)),
             \end{array}
   \right.
\end{equation}
where
$$f(i) = \frac{2Md}{(2k-i-1)i + 2k(d-k+1)} \text{ and }
g(i) = \frac{(2d-2k+i+1)i}{2d}.$$

Using the information flow graph $G$, we can see that there are exactly $k$ points in the tradeoff
curve, or equivalently, $k$ intervals in the threshold function $\alpha^*(d,\gamma)$,
which represent the $k$ newcomers. In the mincut equation, the $k$ terms in the
summation are computed as
the minimum between two parameters: the sum of the weights of
the arcs that we have to cut to isolate the corresponding $v^j_{in}$ from
$S$, and the weight of the arc that we have to cut to isolate the corresponding $v^j_{out}$ from $S$.
Let the first parameter be called the \textit{income} of the corresponding newcomer $s_j$.
Note that the income of the newcomer $s_j$ depends on the previous newcomers.

\subsection{Static cost model}
\label{1costModel}

In \cite{Ak01}, \textit{Akhlaghi et al.} presented another distributed storage
model, where the storage
nodes $V_s$ are partitioned into two sets $V^1$ and $V^2$ with different repair bandwidth. Let
$V^1 \subset V_s$ be the ``cheap bandwidth'' nodes, where each data unit sent
costs
$C_c$, and $V^2 \subset V_s$ be the ``expensive bandwidth'' nodes, where each
data unit sent costs $C_e$ with $C_e > C_c$. This means that when a newcomer
replaces a lost storage node, the cost of downloading data from a node in the
set $V^1$ will be lower than the cost of downloading the same amount of data
from a node in the set $V^2$.

Consider the same situation as in the model described in Subsection \ref{FirstModel}. However, when a storage
node fails, the newcomer node $s_j$, $j=n+1,\ldots,\infty$, connects to
$d_1$ existing storage nodes from $V^1$ sending each one of them $\beta_c$ data units to
$s_j$, and to $d_2$ existing storage nodes from $V^2$ sending each one of them $\beta_e$
data units to $s_j$. Let $d=d_1+d_2$ be the number of helper nodes. Assume that $d$, $d_1$, and $d_2$
are fixed, that is, they do not depend on the storage node $s_j$,
$j=n+1,\ldots,\infty$. In terms of the information flow graph $G$, there is one
arc from $v_{out}^i$ to $v_{in}^j$ of weight $\beta_c$ or $\beta_e$, depending
on
whether $s_i$ sends $\beta_c$ or $\beta_e$ data units, respectively, in
the regenerating process. This new vertex $v_{in}^j$, is also connected to
its
associated $v_{out}^j$ with an arc of weight $\alpha$.

Let the repair cost be $C_T = d_1 C_c \beta_c + d_2 C_e
\beta_e$ and the repair bandwidth $\gamma = d_1 \beta_c + d_2 \beta_e$. To
simplify the model, we can assume, without loss of generality, that $\beta_c =
\tau \beta_e$ for some real number $\tau \ge 1$. This means that we
minimize the repair cost $C_T$ by downloading more data units from the
``cheap bandwidth'' set of nodes $V^1$ than from the ``expensive bandwidth'' set of nodes
$V^2$. Note that if $\tau$ is increased, the repair cost is decreased
and vice-versa. Again it must be satisfied that $\min(\mbox{mincut}(S,DC))\ge
M$.

When $k \le d_1$, the mincut is $\sum_{i=0}^{k-1} \min
\left\lbrace {(d_1\beta_c + d_2 \beta_e - i\beta_c ), \alpha}\right\rbrace \ge
M$, and when $k>d_1$, it is $\sum_{i=0}^{d_1} \min
\left\lbrace {(d_1 \beta_c + d_2 \beta_e - i\beta_c ), \alpha}\right\rbrace  +
\sum_{i=d_1+1}^{k-1} \min\left\lbrace {(d_1 + d_2  - i)\beta_e ,
\alpha}\right\rbrace  \ge M$.
After applying $\beta_c = \tau \beta_e$ and an optimization process,
the mincut equations leads to the threshold function shown in \cite{Ak01}.

\section{Rack model}
\label{rackModel}
In a realistic distributed storage environment, the storage devices are
organized in racks. In this case, the repair cost between nodes which are in the same rack
is much lower than between nodes which are in different racks.

Note the difference of this model compared with the one presented in Subsection
\ref{1costModel}. In that model, there is a static classification of the
storage nodes between ``cheap bandwidth'' and ``expensive bandwidth'' ones. In
our new model, this classification depends on each newcomer. When a storage
node fails and a newcomer enters into the system, nodes from the same rack are
 in the ``cheap bandwidth'' set, while nodes in other racks are in the
``expensive bandwidth'' set. In this paper, we analyze the case
when there are only two racks. Let $V_1$ and $V_2$ be the sets of $n_1$ and $n_2$ storage nodes
from the first and second rack, respectively.

Consider the same situation as in Subsection \ref{1costModel}, but now the sets
of ``cheap bandwidth'' and ``expensive bandwidth'' nodes depend on the
specific replaced node.
Again, we can assume, without loss of generality, that $\beta_c =
\tau \beta_e$ for some real number $\tau \ge 1$. Let the newcomers be the storage nodes $s_j$,
$j=n+1,\ldots, \infty$. Let $d= d_1+d_2$ be the number
of helper nodes for any newcomer, where $d_1$ and $d_2$ are the number of helper nodes
in the first and second rack, respectively.  We can always assume that $d_1 \le d_2$, by swapping
racks if it is necessary.

In both models presented in Section \ref{sec:1}, the repair bandwidth $\gamma$ is the same for
any newcomer. In the rack model, it depends on the rack where the newcomer is placed. Let
$\gamma^1=  \beta_e(d_1 \tau +d_2)$ be the repair bandwidth for any
newcomer in the first rack with repair cost $C_T^1 = \beta_e (C_c d_1 \tau + C_e
d_2)$, and let $\gamma^2= \beta_e (d_2 \tau +d_1)$ be the repair bandwidth for
any newcomer in the second rack with repair cost $C_T^2 = \beta_e (C_c d_2 \tau
+ C_e d_1)$. Note that if $d_1=d_2$ or $\tau=1 $, then $\gamma^1 = \gamma^2$,
otherwise $\gamma^1 < \gamma^2$. To represent a distributed storage system, the
information flow graph is restricted to $\gamma \ge \alpha$ \cite{Di01}. In the
rack model it is a necessary condition that $\gamma^1 \ge \alpha$, which means
that $\gamma^2 \ge \alpha$.

Moreover, unlike the models presented in Section \ref{sec:1}, where it is
straightforward to establish which is the set of nodes which minimize the
mincut,
in the rack model, this set of nodes may change depending on the parameters $k$,
$d_1$, $n_1$ and $\tau$. Recall that the income of a newcomer $s_j$,
$j=n+1,\ldots,\infty$, is the sum of the
weights of the arcs that should be cut in order to isolate $v_{in}^j$ from $S$.
Let $I$ be the indexed multiset containing the incomes of $k$ newcomers which
minimize the mincut. It is easy to see that in the model presented in
Subsection \ref{FirstModel}, $I=\{ (d-i)\beta \;|\; i=0,\ldots, k-1 \}$,
and in the one presented in Subsection \ref{1costModel}, $I=\{ ((d_1-i) \tau
 +d_2) \beta_e \;|\; i=0,\ldots, \min \{d_1, k-1\} \} \cup \{ (d_2-i) \beta_e
\;|\; i=1,\ldots, \min \{d_2, k-d_1-1\} \}$.

In order to establish $I$ in the rack model, the set of $k$ newcomers which
minimize the mincut must be found. First, note that since $d_1 \le d_2$, the
income of the newcomers is minimized by replacing first $d_1$ nodes from the
rack with less number of helper nodes, which in fact minimizes the
mincut. Therefore, the indexed multiset $I$ always contains the incomes of
a set of $d_1$ newcomers from $V_1$. Define $I_1=\{((d_1-i)\tau + d_2)\beta_e \;| \; i=0,\ldots, \min \{d_1,
k-1\} \} $ as the indexed multiset where $I_1[i]$, $i=0,\ldots, \min \{d_1,
k-1\}$, are the incomes of this set of $d_1$ newcomers from $V^1$. If $k-1 \le d_1$,
then $I=I_1$, otherwise $I_1 \subset I$ and $k-d_1-1$
more newcomers which minimize the mincut must be found.

At this point there are two possibilities: either the remaining nodes from $V_1$ are
in the set of newcomers which minimize the mincut or not. Define $I_2 = \{ d_2
\beta_e \;|\; i=1,\ldots, \min \{k-d_1-1, n_1-d_1-1 \}\} \cup \{ (d_2-i) \tau
\beta_e \;|\; i = 1,\ldots, \min \{d_2, k-n_1 \} \}$ as the indexed multiset
where $I_2[i]$, $i=0,\ldots, k-d_1-2$, are the incomes of a set of $k-d_1-1$ newcomers,
including the remaining $n_1-d_1-1$ newcomers from $V_1$ and newcomers from $V_2$. Note that if
$n_1-d_1-1 > k-d_1-1$, it only contains newcomers from $V_1$. Define $I_3
= \{ (d_2-i) \tau \beta_e \; |\;i=1,\ldots, \min \{d_2, k-d_1-1\} \}$ as the
indexed multiset where $I_3[i]$, $i=0,\ldots, k-d_1-2$, are the incomes of a
set of $k-d_1-1$ newcomers from $V_2$. Note that when $i > d_2$ in $I_2$ or
$I_3$ the resulting income is negative, which is not possible. In fact, given by
the information flow graph, the income for any further newcomer is zero. It can
be assumed that $d_2 \ge k-d_1-1 \ge k-n_1$, because the mincut equation does
not change when $d_2 < k-d_1-1$ or $d_2<k-n_1$.

\begin{prop}
\label{prop:01}
As $|I_2| = |I_3| = k-d_1-1$, if $\sum_{i=0}^{k-d_1-2} I_2[i] <
\sum_{i=0}^{k-d_1-2} I_3[i]$, then $I=I_1 \cup
I_2$; and if $\sum_{i=0}^{k-d_1-2} I_2[i] \ge \sum_{i=0}^{k-d_1-2} I_3[i]$, then
$I=I_1 \cup I_3$.
\end{prop}
\begin{demo}
Let $J$ be an indexed multiset containing the incomes of a set
of newcomers such that $I = I_1 \cup J$. It can be seen that
either $J = I_2$ or $J = I_3$.
\end{demo}

By using Proposition \ref{prop:01}, if $I =
I_1 \cup I_2$, the corresponding mincut equation is $\sum_{i=0}^{|I_1|-1} \\ \min
\left\lbrace I_1[i], \alpha \right\rbrace + \sum_{i=0}^{|I_2|-1} \min
\left\lbrace I_2[i], \alpha\right\rbrace \ge M$; and if $I = I_1 \cup I_3$, the
equation is $\sum_{i=0}^{|I_1|-1} \\ \min \left\lbrace I_1[i], \alpha \right\rbrace
+ \sum_{i=0}^{|I_3|-1} \min \left\lbrace I_3[i], \alpha\right\rbrace \ge M$.

In the previous models, described in Section \ref{sec:1}, the decreasing behavior of the incomes
included in the mincut equation is used to find the threshold function to minimize the parameters
$\alpha$ and $\gamma$.  In the rack model, the incomes in the mincut equations may not have a
decreasing behavior as the newcomers enter into the system. Therefore, it
is not possible to find the threshold function as it is done in the previous
models. However, we give a threshold function for the rack model described in
this section, which represents the behavior of the mincut equations also for the
previous models. Note that the way to represent this threshold function can be
seen as a generalization, since it also represents the behavior for the previous
given models.   

Let $L$ be the increasing ordered list of values such that for all $i, \; i=
0,\ldots, k-1$, $I[i]/\beta_e \in L$ and $|I|=|L|$.
Note that any of the information flow graphs representing any model from Section
\ref{sec:1} and any of the ones representing the rack model, can be described in
terms of $I$, so they can be represented by $L$. Therefore, once $L$ is found,
it is possible to find the parameters $\alpha$ and $\beta_e$ (and then $\gamma$
or $\gamma^i$, $i=1,2$) using the following threshold function.

\begin{theo}
 The threshold function $\alpha^*(d_1,d_2,\beta_e)$ (which also depends on
$\tau$ and $k$) is the following:
\begin{equation}
\label{treshold3}
 \alpha^*(d_1,d_2,\beta_e) = \left\{ \begin{array}{ll}
             \frac{M}{k}, & \beta_e \in [f(0), +\infty)  \\ & \\
              \frac{M-g(i)\beta_e}{k-i}, & \beta_e \in
[f(i),f(i-1))\\ &
i=1, \ldots, k-1,
   \end{array}
   \right.
\end{equation}
subject to $\gamma^1=(d_1 \tau + d_2) \beta_e \ge \alpha$,
where
$$f(i) = \frac{M}{L[i](k-i)+g(i)} \text{ and }
g(i) =\sum_{j=0}^{i-1} L[j].$$
\end{theo}

It can happen that two values in $L$ are equal, so $f(i) = f(i-1)$. In this
case, we consider that the interval $[f(i),f(i-1))$ is empty.
Note that the threshold function (\ref{treshold3}) is subject to $\gamma^1=(d_1
\tau + d_2) \beta_e \ge \alpha$. However, $\gamma^1 \ge \alpha$ is only
satisfied when the highest value of $I_1$ divided by $\beta_e$ coincides with the highest value of
$L$. By definition, $\max I_1  = I_1[0]$, so $\max L = I_1[0]/\beta_e$. In
terms of the tradeoff curve, this means that there is no point in the curve that
outperforms the MBR point. In order to achieve that $\gamma_1 \ge \alpha$, it is necessary
that $f(i) \ge \frac{M}{\frac{I_1[0]}{\beta_e} (k-i) + g(i)}$ for
$i=0,\ldots,k-1$. This restriction is achieved by removing from
$L$ any value $L[i]$ such that $L[i] > I_1[0]/\beta_e$, $i=0,\ldots,k-1$. From
now on, we assume that $L[|L|-1] = I_1[0]/\beta_e$.

\begin{figure}
\centering
\begin{tikzpicture}[shorten >=1pt,->,font= \small]
  \tikzstyle{vertix}=[circle,fill=black!25,minimum size=18pt,inner
sep=0pt,font=\tiny]
  \tikzstyle{invi}=[circle]
  \tikzstyle{background}=[rectangle, fill=gray!20, inner sep=0.2cm,rounded
corners=5mm]

  \node[vertix] (s) {$S$};

  \node[vertix, right= 1cm of s] (vin_4)  {};

  \node[vertix, above=0.6cm of vin_4] (vin_3)  {};
  \node[vertix, below=0.05 of vin_4] (vin_5)  {};

  \node[vertix, above=0.05 of vin_3] (vin_2)  {$v_{in}^2$};
  \node[vertix, below=0.05 of vin_5] (vin_6)  {$v_{in}^n$};

  \node[vertix,above=0.05 of vin_2] (vin_1)  {$v_{in}^1$};

 \foreach \to in {1,2,3}
   {\path (s) edge[bend left=20, font= \tiny] node[anchor=south,above]{$\infty$}
(vin_\to);}
 \foreach \to in {4,5,6}
   {\path (s) edge[bend right=20, font= \tiny]
node[anchor=south,above]{$\infty$}
 (vin_\to);}

  \node[vertix, right= 0.5cm of vin_1] (vout_1)  {$v_{out}^1$};
  \node[vertix, right= 0.5cm of vin_2] (vout_2)  {$v_{out}^2$};
  \node[vertix, right= 0.5cm of vin_3] (vout_3)  {};
  \node[vertix, right= 0.5cm of vin_4] (vout_4)  {};
  \node[vertix, right= 0.5cm of vin_5] (vout_5)  {};
  \node[vertix, right= 0.5cm of vin_6] (vout_6)  {$v_{out}^n$};

  \node[vertix, right= 1cm of vout_1] (vin_7)
{\scriptsize{$v_{in}^{n+1}$}};
  \node[vertix, right= 0.5 cm of vin_7] (vout_7)
{\scriptsize{$v_{out}^{n+1}$}};

  \node[vertix, below= 0.5cm of vout_7]
(vin_8)
{\scriptsize{$v_{in}^{n+2}$}};
  \node[vertix, right= 1cm of vin_8] (vout_8)
{\scriptsize{$v_{out}^{n+2}$}};

  \node[vertix, below= 0.5cm of vout_8]
(vin_9)
{\scriptsize{$v_{in}^{n+3}$}};
  \node[vertix, right=0.5cm of vin_9] (vout_9)
{\scriptsize{$v_{out}^{n+3}$}};

  \node[vertix, below=0.5cm of vout_9]
(vin_10)
{\scriptsize{$v_{in}^{n+4}$}};
  \node[vertix, right=0.5cm of vin_10] (vout_10)
{\scriptsize{$v_{out}^{n+4}$}};

  \node[vertix, right=2.5cm of vout_9] (DC)
{DC};

 \foreach \to in {7,8,9,10}
   {\path (vout_\to) edge[bend left=20, font= \tiny]
node[anchor=south,above]{$\infty$}
 (DC);}

  \path[->, bend left, font= \tiny] (vout_2) edge
node[anchor=south,above]{$\beta_c$} (vin_7);
  \path[dashed,->, bend right,font= \tiny] (vout_4) edge
node[anchor=south,above]{$\beta_e$}(vin_7);
  \path[->, dashed, bend right,font= \tiny] (vout_5) edge
node[anchor=south,above]{$\beta_e$} (vin_7);
  \path[->, dashed, bend right,font= \tiny] (vout_6) edge
node[anchor=south,above]{$\beta_e$} (vin_7);

  \path[->] (vout_7) edge node[anchor=south,left,font= \tiny]{$\beta_c$}
(vin_8);
  \path[->, dashed, bend right,font= \tiny] (vout_4) edge
node[anchor=south,above]{$\beta_e$} (vin_8);
  \path[->, dashed, bend right,font= \tiny] (vout_5) edge
node[anchor=south,above]{$\beta_e$} (vin_8);
  \path[->, dashed, bend right,font= \tiny] (vout_6) edge
node[anchor=south,above]{$\beta_e$} (vin_8);

  \path[->, dashed, bend left=65,font= \tiny] (vout_7) edge node[anchor=south,
above]{$\beta_e$}
(vin_9);
  \path[->,dashed,font= \tiny] (vout_8) edge node[anchor=south,left]{$\beta_e$}
(vin_9);
  \path[->, bend right=20,font= \tiny] (vout_5) edge
node[anchor=south,above]{$\beta_c$} (vin_9);
  \path[->, bend right=20,font= \tiny] (vout_6) edge
node[anchor=south,above]{$\beta_c$} (vin_9);

  \path[->, bend left=65, dashed, font= \tiny] (vout_7) edge node[anchor=south,
above]{$\beta_e$}
(vin_10);
  \path[->, bend left=65, dashed, font= \tiny] (vout_8) edge
node[anchor=south,above]{$\beta_e$}
(vin_10);
  \path[->] (vout_9) edge node[anchor=south,left,font= \tiny]{$\beta_c$}
(vin_10);
  \path[->, bend right=10, font= \tiny] (vout_6) edge
node[anchor=south,above]{$\beta_c$} (vin_10);

 \foreach \from/\to in {1,2,3,4,5,6,7,8,9,10}
  { \path[->, font= \tiny] (vin_\from) edge node[anchor=south] {$\alpha$}
(vout_\to); }

\begin{pgfonlayer}{background}
 \node [background,fit= (vin_1) (vin_3) (vout_3)] {};
 \node [background,fit= (vin_4) (vin_6) (vout_6)] {};
\end{pgfonlayer}
\end{tikzpicture}
\caption{ \small Information flow graph corresponding to the rack model
when $k > d_1$,
with $k=4$, $d_1=1$, $d_2=3$, and $n_1=n_2=3$.
}
\label{Figure:2}
\end{figure}
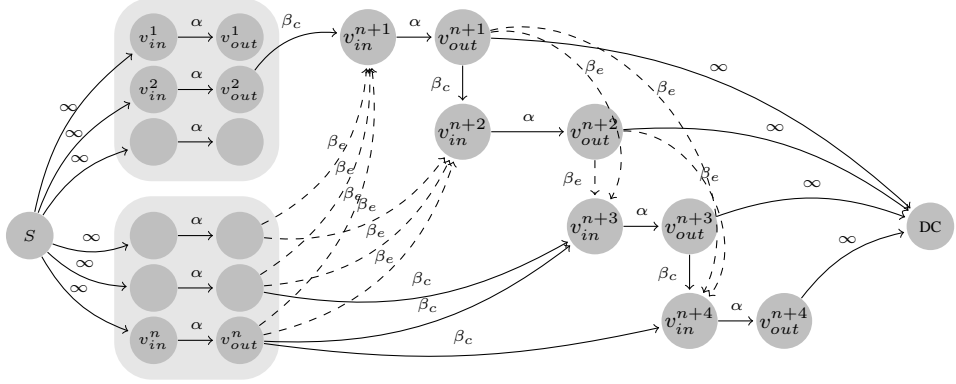

When $k \le d_1$, the mincut equations and the threshold function
(\ref{treshold3}) of the rack model are exactly the same as the ones
shown in \cite{Ak01} for the model described in Subsection \ref{1costModel}.
Indeed, it can be seen that when $k \le d_1$, the rack model and the static cost
model have the same behavior because $I=I_1$.

Figure \ref{Figure:2} shows the example of an information flow graph corresponding
to a regenerating code with $k=4$, $d_1=1$, $d_2=3$, and $n_1=n_2=3$. Taking for
example $\tau= 2$, we have that $I_1=\{5\beta_e,3\beta_e \}$, $I_2=
\{3\beta_e,4 \beta_e \}$ and $I_3=\{4\beta_e,2\beta_e \}$. By Proposition
\ref{prop:01}, since $\sum_{i=0}^{1} I_2[i] > \sum_{i = 0}^{1}I_3[i]$, $I= I_1
\cup I_3 = \{5\beta_e,3\beta_e,4\beta_e,2\beta_e\}$, and then $L=[2,3,4,5]$.
Applying the corresponding mincut equation to the threshold function
(\ref{treshold3}), we have that
\begin{equation}
 \alpha^*(d_1,d_2,\beta_e) = \left\{ \begin{array}{ll}
             \frac{M}{4}, & \beta_e \in [\frac{M}{8}, +\infty)  \\ &\\
             \frac{M-2 \beta_e}{3}, & \beta_e \in
[\frac{M}{11},\frac{M}{8})\\ & \\
	       \frac{M-5 \beta_e}{2}, & \beta_e \in
[\frac{M}{13},\frac{M}{11})\\ & \\
	       M-9 \beta_e, & \beta_e \in [\frac{M}{14},\frac{M}{13}).\\
   \end{array}
   \right.
\end{equation}

\subsection*{MSR and MBR points}

The threshold function (\ref{treshold3}) leads to a tradeoff curve
between $\alpha$ and $\beta_e$. Note that, like in the static cost model,
since there is a different repair bandwidth $\gamma_1$ and $\gamma_2$ for each rack,
this curve is based on $\beta_e$ instead of $\gamma_1$ and $\gamma_2$.

At the MSR point, the amount of stored data per node is $\alpha_{MSR} = M/k$.
Moreover, at this point, the minimum value of $\beta_e$ is $\beta_e = f(0) =
\frac{M}{L[0]k}$, which leads to $$\gamma^1_{MSR} = \frac{(d_1 \tau +
d_2)M}{L[0]k} \quad \textrm{and} \quad  \gamma^2_{MSR} = \frac{(d_2 \tau + d_1)M}{L[0]k}.$$
On the other hand, at the MBR point, as $f(i)$ is a
decreasing function, the parameter $\beta_e$ which leads to the minimum
repair bandwidths is $\beta_e = f(|L|-1) = \frac{M}{L[|L|-1](k-|L|+1)+g(|L|-1)} $. Then,
the corresponding amount of stored data per node is $\alpha_{MBR} =
\frac{M (L[|L|-1]k-|L|+1)}{(k-|L|+1)^2 L[|L|-1] + g(|L|-1)}$, and the repair bandwidths
are $$\gamma^1_{MBR} = \frac{(d_1 \tau + d_2)M}{L[|L|-1](k-|L|+1)+g(|L|-1)} \quad \textrm{and}$$
$$\gamma^2_{MBR} = \frac{(d_2 \tau + d_1)M}{L[|L|-1](k-|L|+1)+g(|L|-1)}.$$

\section{Analysis}
\label{sec:3}

\begin{figure}
\centering
\begin{tabular}[h]{cc}
\begin{tikzpicture}
 \begin{axis}
[ xlabel=$\gamma$,
  ylabel=$\alpha$,
  xmax= 0.58,
  ymax=0.3,
  height=3.5cm,
  width=6cm,
  legend style={font=\tiny},
]

\addplot[color=red, mark=square] coordinates { (0.6,0.2000000000)
(0.3000000000,0.2000000000)
(0.2727272727,0.2045454545) (0.2553191489,0.2127659574)
(0.2448979592,0.2244897959) (0.2400000000,0.2400000000) (0.2400000000, 0.4) };
%\addlegendentry{$\tau=1$}

\addplot[color=black,mark=*] coordinates { (0.6,0.2000000000)
(0.3600000000,0.2000000000)
(0.3103448276,0.2068965517) (0.2812500000,0.2187500000)
(0.2647058824,0.2352941176) (0.2571428571,0.2571428571) (0.2571428571, 0.4) };
%\addlegendentry{$\tau=2$}

\addplot[color=blue,mark=x] coordinates { (0.6,0.2000000000)
(0.4500000000,0.2000000000)
(0.3600000000,0.2100000000) (0.3130434783,0.2260869565)
(0.2880000000,0.2480000000) (0.2769230769,0.2769230769)  (0.2769230769, 0.4)};
%\addlegendentry{$\tau=5$}

\addplot[color=green,mark=triangle] coordinates { (0.6,0.2000000000)
 (0.5076923077,0.2000000000)
(0.3882352941,0.2117647059) (0.3300000000,0.2300000000)
(0.3000000000,0.2545454545) (0.2869565217,0.2869565217) (0.2869565217, 0.4)};
%\addlegendentry{$\tau=10$}
\legend{$\tau=1$, $\tau=2$, $\tau=5$, $\tau=10$}
\end{axis}
\end{tikzpicture}
&
\begin{tikzpicture}
 \begin{axis}
[ xlabel=$\beta_e$,
  ylabel=$\alpha$,
  xmax= 0.055,
  ymax=0.18,
  height=3.5cm,
  width=8cm,
  scaled x ticks = false,
  legend style={font=\tiny},
  x tick label style={/pgf/number format/fixed, /pgf/number format/1000 sep =
\thinspace}
]

\addplot[color=red,mark=square] coordinates { (1.1,
0.1)(0.05000000000,0.1000000000)
(0.03448275862,0.1034482759) (0.02702702703,0.1081081081)
(0.02272727273,0.1136363636) (0.02000000000,0.1200000000)
(0.01818181818,0.1272727273) (0.01694915254,0.1355932203)
(0.01612903226,0.1451612903) (0.01562500000,0.1562500000)
(0.01538461538,0.1692307692) (0.01538461538, 0.3)};
%\addlegendentry{$\tau=1$}

\addplot[color=black,mark=*] coordinates { (1.1,0.1)
(0.04166666667,0.1000000000) (0.02873563218,0.1034482759)
(0.02252252252,0.1081081081) (0.01893939394,0.1136363636)
(0.01700680272,0.1224489796) (0.01572327044,0.1320754717)
(0.01488095238,0.1428571429)(0.01436781609,0.1551724138)
(0.01412429379,0.1694915254) (0.01412429379,0.4)};
%\addlegendentry{$\tau=6/5$}

\addplot[color=blue,mark=x] coordinates { (1.1,0.1) (0.02500000000,0.1000000000)
(0.01724137931,0.1034482759)
(0.01388888889,0.1111111111) (0.01219512195,0.1219512195)
(0.01136363636,0.1363636364) (0.01086956522,0.1521739130)
(0.01063829787,0.1702127660) (0.01063829787, 0.4)};
%\addlegendentry{$\tau=2$}

\addplot[color=green,mark=triangle] coordinates { (1.1, 0.1)
(0.01666666667,0.1000000000) (0.006666666667,0.1066666667)
(0.005494505494,0.1098901099) (0.004464285714,0.1160714286)
(0.004032258065,0.1209677419) (0.003597122302,0.1294964029)
(0.003401360544,0.1360544218) (0.003205128205,0.1474358974)
(0.003125000000,0.1562500000) (0.003067484663,0.1717791411)
(0.003067484663,0.4)};
%\addlegendentry{$\tau=10$}

\legend{$\tau=1$, $\tau=6/5$,$\tau=2$, $\tau=10$}
\end{axis}
\end{tikzpicture}
\end{tabular}
\caption{ \small \label{plots:1} Left: tradeoff curves between
$\alpha$ and $\gamma$ for $k=5$, $d_1=6$, $d_2=6$, and $M=1$, so $k \le d_1$.
Right: tradeoff curves between $\alpha$ and $\beta_e$ for $k=10$, $d_1=5$,
$d_2=6$, $n_1=n_2=6$,
and $M=1$, so $k>d_1$.}
\end{figure}
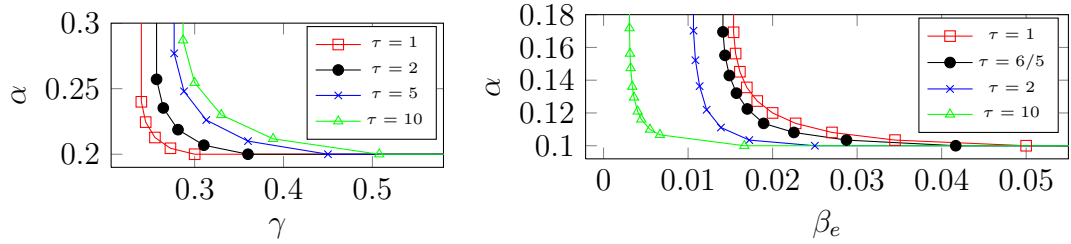

In this section, we analyze the results of the new fundamental
tradeoff curve shown in Section \ref{rackModel} for the rack model. We also  compare these
results with previous contributions of papers \cite{Di01} and \cite{Ak01}
provided it can be carried out.

When $\tau=1$, we have that $\beta_e = \beta_c$, so $\gamma = d \beta_e$.
This corresponds to the same case as in the fundamental tradeoff curve shown in
Subsection \ref{FirstModel}, since one can assume that $\beta_e = \beta$.
When $\tau > 1$ and $k \le d_1$, the rack model coincides with the one
presented in Subsection \ref{1costModel} and it uses more repair bandwidth than
the one
shown in Subsection \ref{FirstModel} as it is explained in \cite{Ak01}. Figure
\ref{plots:1} left shows the tradeoff curves  between $\alpha$ and $\gamma$ for the rack model when $k \le
d_1$ (for different values of $\tau$). Note that
as $\tau$ increases, both $\alpha$ and $\gamma$ also increase, but the repair cost 
decreases as further we see in this section. Moreover, both extremal points for each
curve are shown: the MSR point is when $\alpha$ is minimum and the MBR point is
when $\gamma$ is minimum. On the other hand, the case when $\tau > 1$ and $k > d_1$ 
is different from the previous models. An example is shown in Figure \ref{plots:1} right.
Note that as $\tau$ increases $\beta_e$ decreases.

Despite the repair bandwidths $\gamma_1$ and $\gamma_2$ may increase with $\tau$, the repair cost
always decreases. The rack model has two repair bandwidths, $\gamma^1$ and
$\gamma^2$,
this means that it also has two repair costs $C_T^1= \beta_e(C_c d_1 \tau +
C_e d_2)$ and  $C_T^2= \beta_e(C_c d_2 \tau + C_e d_1)$.
As we have said, the case when $\tau=1$ is exactly the same as the one presented
in \cite{Di01}. In this case, for each $i=0,\ldots,k-1$, taking $\gamma=f(i)$, 
we have that $\beta = f(i)/d$.
Then, we can say that $ C_T^1(\tau=1) = \frac{f(i)}{d} (C_c d_1 \tau + C_e d_2)$ and
$ C_T^2(\tau=1) = \frac{f(i)}{d} (C_c d_2 \tau + C_e d_1)$. From
(\ref{thresholdDim}), we know that $f(i) =  \frac{2Md}{(2k-i-1) i
+2k(d_1+d_2-k+1)}$, so finally $C_T^1(\tau=1)= \frac{2Md (C_c d_1 \tau + C_e
d_2)}{(2k-i-1) i +2k(d_1+d_2-k+1)}$ and $C_T^2(\tau=1)= \frac{2Md (C_c d_2 \tau
+ C_e d_1)}{(2k-i-1) i +2k(d_1+d_2-k+1)}$. When $\tau >1$, we have that $\beta_e = f(i)$,
so $C_T^1(\tau>1)= \frac{M(C_c d_1 \tau + C_e d_2)}{L[i](k-i)+g(i)}$ and $C_T^2(\tau>1)=
\frac{M(C_c d_2 \tau + C_e d_1)}{L[i](k-i)+g(i)}$.

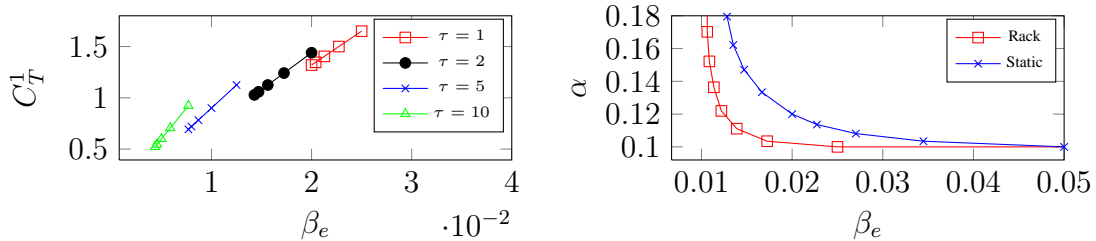
\begin{figure}[h]
\centering
\begin{tabular}[h]{cc}

\begin{tikzpicture}
 \begin{axis}
[ xlabel=$\beta_e$,
  ylabel=$C_T^1$,
  xmax= 0.04,
  ymax=1.8,
  height=3.5cm,
  width=6.8cm,
  legend style={font=\tiny},
  x tick label style={/pgf/number format/fixed, /pgf/number format/1000 sep =
\thinspace}
]

\addplot[color=red, mark=square] coordinates {
(0.02500000000,1.650000000)
(0.02272727273,1.500000000) (0.02127659574,1.404255319)
(0.02040816327,1.346938776) (0.02000000000,1.320000000)};
%\addlegendentry{$\tau=1$}

\addplot[color=black,mark=*] coordinates {
(0.02000000000,1.440000000)
(0.01724137931,1.241379310) (0.01562500000,1.125000000)
(0.01470588235,1.058823529) (0.01428571429,1.028571429) };
%\addlegendentry{$\tau=2$}

\addplot[color=blue,mark=x] coordinates {
(0.01250000000,1.125000000)
(0.01000000000,0.9000000000) (0.008695652174,0.7826086957)
(0.008000000000,0.7200000000) (0.007692307692,0.6923076923)};
%\addlegendentry{$\tau=5$}

\addplot[color=green,mark=triangle] coordinates {
 (0.007692307692,0.9230769231)
(0.005882352941,0.7058823529) (0.005000000000,0.6000000000)
(0.004545454545,0.5454545455) (0.004347826087,0.5217391304)};
%\addlegendentry{$\tau=10$}

\legend{$\tau=1$, $\tau=2$, $\tau=5$, $\tau=10$}
\end{axis}
\end{tikzpicture}

&

\begin{tikzpicture}
 \begin{axis}
[ xlabel=$\beta_e$,
  ylabel=$\alpha$,
  xmax= 0.05,
  ymax=0.18,
  height=3.5cm,
  width=6.8cm,
  scaled x ticks = false,
  legend style={font=\tiny},
  x tick label style={/pgf/number format/fixed, /pgf/number format/1000 sep =
\thinspace}
]

\addplot[color=red, mark= square] coordinates {(1.1,0.1)
(0.02500000000,0.1000000000)
(0.01724137931,0.1034482759)
(0.01388888889,0.1111111111) (0.01219512195,0.1219512195)
(0.01136363636,0.1363636364) (0.01086956522,0.1521739130)
(0.01063829787,0.1702127660) (0.01063829787, 0.4)};
%\addlegendentry{Rack}

\addplot[color=blue,mark=x] coordinates { (1.1, 0.1)
(0.05000000000,0.1000000000) (0.03448275862,0.1034482759)
(0.02702702703,0.1081081081) (0.02272727273,0.1136363636)
(0.02000000000,0.1200000000) (0.01666666667,0.1333333333)
(0.01470588235,0.1470588235) (0.01351351351,0.1621621622)
(0.01282051282,0.1794871795) (0.01250000000,0.2000000000)
(0.01250000000, 0.3)};
%\addlegendentry{Static}
\legend{Rack, Static}
\end{axis}
\end{tikzpicture}

\end{tabular}

\caption{\small\label{plots:2} Left: chart showing the repair cost of
the rack model for $M=1$, $k=5$, $d_1=6$, $d_2=6$, $C_c = 1$ and
$C_e=10$. The points correspond to the $k=5$ values given by
$f(i)$, $i=0,\ldots,4$. Right: chart comparing the rack model
presented in this paper with the static cost model presented in \cite{Ak01} for
$M=1$, $k=10$, $d_1=5$, $d_2=6$, $n_1=n_2=6$ and $\tau=2$. }
\end{figure}

Define $\eta(\tau)=\frac{C_T^1(\tau>1)}{C_T^1(\tau=1)}=\frac{C_T^2(\tau>1)}{C_T^2(\tau=1)}$.
We know that $\beta_e = f(i) =
\frac{M}{L[i](k-i)+g(i)}$, so
$$\eta(\tau) = \frac{ (2k-i-1) i +2k(d_1+d_2-k+1)}
{2d ( L[i](k-i)+g(i) )} $$  is a decreasing function over $\tau$ for every
fixed $i$. This means that as $\tau$ increases, the repair costs $C_T^1$ and $C_T^2$
always decrease. Figure \ref{plots:2} left shows the decreasing behavior
of $C_T^1$ and $\beta_e$ as $\tau$ increases.

When $k\le d_1$, the static cost model and the rack model have the same behavior. However, when
$k>d_1$, it can be seen in Figure \ref{plots:2} right that the rack model
outperforms the static cost model in terms of $\beta_e$ and $\alpha$. Note that the
repair cost $C_T$ of the static model is equivalent to $C_T^1$ of the rack model.
Fixed $d_1$, $d_2$, and $\tau$, as  $\beta_e$ decreases $C_T^1$ does,
so we can say that the rack model also outperforms the static cost model in terms of
repair cost. In \cite{Ak01}, the authors show that the static cost model outperforms the
basic model presented in \cite{Di01} in terms of repair cost. Therefore, it comes
straightforward that the rack model
also outperforms the basic model in terms of repair cost.

\section{Conclusions}
\label{sec:4}

In this paper, a new mathematical model for a distributed storage environment is
presented and analyzed. In this new model, the cost of downloading data units
from nodes in different racks is introduced. That is, the cost of downloading
data units from nodes located in the same rack is much lower than the cost of
downloading data units from nodes located in a different rack. The rack model
is an approach to a more realistic distributed storage environment like the ones
used in companies dedicated to the task of storing information over a network.

The rack model is deeply analyzed in the case that there are two racks. The
differences between this model and previous models are shown. Due to it is a
less simplified model compared to the ones presented previously, the rack model
introduces more difficulties in order to be analyzed. In this paper, we provide
a complete analysis of the model including some
important contributions like the generalization of the process to find the
threshold function of a distributed storage system. This new threshold function
fits in any previous model and allows to represent the information flow graphs considering
different repair costs.

We provide the general threshold function and apply it to the model when there
are two racks. We provide the tradeoff curve between the repair bandwidth and
the amount of stored data per node and compare it to the ones found in previous
models. We also analyze the repair cost of this new model, and we  conclude that
the rack model outperforms previous models in terms of repair cost.

\bibliographystyle{IEEEtran}
\bibliography{IEEEabrv,references}

\end{document}